\documentstyle[epsf]{ioplppt}        

\newcommand{\beqn}{\begin{equation}}
\newcommand{\eeqn}{\end{equation}}
\newcommand{\pu}{\qquad .}

\newcommand{\rcl}{R_{\rm cl}}
\newcommand{\ncl}{N_{\rm cl}}
\newcommand{\rhc}{R_{\rm hc}}
\newcommand{\lcl}{\lambda_{\rm cl}}
\newcommand{\lhc}{\lambda_{\rm hc}}
\newcommand{\lpar}{\lambda_{\rm par}}
\newcommand{\phc}{p_{\rm hc}}
\newcommand{\chia}{\chi_{\rm A}}
\newcommand{\cova}{C_{\rm A}}
\begin{document}
\jl{13}

\title{Computer simulation of crystallization kinetics with non-Poisson
distributed nuclei}[Simulation of crystallization kinetics]

\author{Patric Uebele and Helmut Hermann} 

\address{Institut f\"ur Festk\"orper- und Werkstofforschung Dresden, PF
27 00 16, D-01171 Dresden, Germany} 

\begin{abstract}
The influence of non-uniform distribution of nuclei on crystallization
kinetics of amorphous materials is investigated. This case cannot be
described by the well-known Johnson-Mehl-Avrami (JMA) equation, which is
only valid under the assumption of a spatially homogeneous nucleation
probability. The results of computer simulations of crystallization
kinetics with nuclei distributed according to a cluster and a hardcore
distribution are compared with JMA kinetics. The effects of the
different distributions on the so-called Avrami exponent $n$ are shown.\\
Furthermore, we calculate the small-angle scattering curves of the
simulated structures which can be used to distinguish experimentally 
between the three nucleation models under consideration.
\end{abstract}

%
%

\section{\label{intro}Introduction}
The properties of metallic glasses and other amorphous materials may be
impaired by even small amounts of crystalline phases. Crystallization
can also improve the properties of some amorphous materials, e.g.~glass
ceramics. So the understanding of
the crystallization process is very important. The analysis of
experimental data is often made within the framework of the
Avrami theory \cite{avrami39,johnson39} by means of the
Johnson-Mehl-Avrami equation, which gives a relation between the
fraction of transformed (i.e.~crystallized) material $\chi (t)$ and the time at
constant temperature. An equivalent approach was made by Kolmogorov
\cite{shepilov94}.\\
Both models are based on the following assumptions:
\begin{enumerate}
\item \label{1} Crystallization is considered in an unlimited medium.
\item \label{2} Nucleation of crystals begins at time $t=0$ and occurs in a
non-crystallized region. The nucleation rate per unit volume, $\alpha
(t)$, is assumed to be independent of the coordinates.
\item \label{3} The growth of crystals ceases at the points of mutual
impingements, whereas it continues unchanged elsewhere. Before they
touch each other, the crystals have a geometrically similar and convex shape
(the Avrami approach is restricted to spherical crystals).
\item \label{4} The growth rate $v(t)$ in a given direction is the same for all
crystals and depends only on time.\\
\end{enumerate}
Based on these assumptions, one can derive the following exact 
relation:
\begin{equation}
\label{JMA1}	
\chi (t) = 1 - \exp{\left [ -\int\limits_0^t \alpha(\tau) V(\tau,t) \mbox{d}\tau
\right ] }
\end{equation}
where 
\begin{equation}
V(\tau,t) = V_0 \left [\int\limits_{\tau}^{t} v(t^{\prime}) \mbox{d}
t^{\prime} \right ]^3 \pu
\end{equation}
$V_0$ is a form factor, in the case of spherical crystals $V_0 = 4\pi/3$.
An integration of equation (\ref{JMA1}) is only possible by making
specific assumptions about the time dependence of the nucleation rate
$\alpha(t)$. \\
If both $v$ and $\alpha$ are independent of time (interface controlled
growth and continuous nucleation), then
\beqn
\label{JMA2}
\chi(t)=1-\exp\left [-\frac{1}{4}V_0 \alpha v^3 t^4 \right ] \pu
\eeqn
If only the growth rate $v$ is constant and all
crystals are formed 
simultaneously at $t=0$ with a mean number density $\beta$
(instantaneous nucleation), the 
resulting nucleation rate $\alpha(t)=\beta \delta(t)$ can be substituted
into equation (\ref{JMA1}). In this case
\beqn
\label{JMA3}
\chi(t) = 1 - \exp \left [ - \beta V_0 v^3 t^3 \right ] \pu
\eeqn
Avrami proposed that for a three-dimensional nucleation and
growth process with constant or decreasing nucleation rate, the general
relation
\beqn
\label{JMA4}
\chi(t) = 1 - \exp(-k t^n)
\eeqn
should describe the crystallized volume fraction.
Equation (\ref{JMA4}) is the so-called Avrami equation with the Avrami
exponent $n$, where $3\le n \le 4$.\\
This equation is often used to analyse experimental data by means of
a logarithmic plot, where $\ln \left \{ - \ln [1- \chi(t) ] \right \}$ is
plotted versus $\ln(t)$. The slope of the resulting straight line is the
Avrami exponent $n$, which describes crystallization kinetics. But
if the previously mentioned assumptions are not exactly satisfied, the
resulting Avrami exponents $n$ may be misleading. We study cases
where some of the assumptions (\ref{1}-\ref{4}) are violated and no
analytical results for the crystallization kinetics are available.\\
If the chemical composition of the two phases involved in the
transformation is different, the growth
rate is diffusion controlled. In this case, the radii $r$ of spherical
crystals grow according to 
\beqn
\label{difflaw}
r(t^{\prime},t) = g \cdot \sqrt{t - t^{\prime}} 
\eeqn
where $g$ is a constant and $t$ and $t^{\prime}$ are the observation and
nucleation times, respectively.
Substituting (\ref{difflaw}) into equation (\ref{JMA1}) leads to an
Avrami-exponent $n$ with $1.5 \le n \le 2.5$. But if continuous
nucleation occurs, assumption (\ref{4}) is violated, as the growth rate
decreases with increasing life time of the crystal. As a consequence,
crystals nucleated at different times have different growth rates. In
the Johnson-Mehl method, the growth law (\ref{difflaw}) allows ``phantom
crystals" (crystals that are nucleated in an already crystallized area)
to outgrow the real crystal. Therefore, the crystallized volume
fraction is overestimated.\\
The assumption (\ref{2}) has no physical reasons. In practice,
nucleation may occur preferentially in certain macroscopic regions
(e.g.~grain boundaries), so that the nucleation probability becomes
dependent on the coordinates \cite{christian}.\\
Furthermore, the grains have to reach a 
critical size $r_{\rm crit}$ to start growing.This critical radius is
determined by the free energy and the surface 
energy. Therefore, in a sphere of radius $r_{\rm crit }$ only one crystal can
nucleate \cite{christian}. This also violates assumption (\ref{2}).\\
In the present paper the influence of non-uniformly distributed nuclei
and critical radii on crystallization kinetics is studied by computer
simulation. We use methods of stochastic geometry
\cite{stoyan83,stoyan87,hermann91}, namely the germ-grain models. The
models are explained in section \ref{models} and the simulation technique
is pointed out in section \ref{simulation}. The results are presented in
section \ref{results}.

\section{\label{models}The models}
A germ-grain model is defined by a point field $P_1,
P_2, ...$ with density $\lambda$ and a series of grains $A_1, A_2,...$
with finite size. The complete model $A$ is formed by the union of the
grains $A_n$ shifted to the points $P_n$. Due to the
diversity of point 
fields and types of possible grains there is a great variety of
germ-grain models. In our study, we always use spherical grains of
radius $r_i$. \\
To model the pure JMA case with assumptions (\ref{1}-\ref{4}) fulfilled,
we use a {\em Poisson model} for the underlying 
point field. The two fundamental properties of this point field are:
\begin{itemize}
\label{poisson}
\item The number, $N(G)$, of points lying in an arbitrary region $G$
with volume $V(G)$ is a random variable. The probability $P$ of finding
$n$ points in the region $G$ is given by the Poisson distribution
\beqn
P(N(G)=n) = \frac{ [ \lambda V(G) ]^n}{n!} \exp[- \lambda V(G)] 
 \quad n=1,2,... \pu
\eeqn
\item Considering disconnected regions $G_1, G_2, ...$, the numbers
$N(G_1), N(G_2), ...$ are independent random variables. 
\end{itemize}
Germ-grain models with an underlying Poisson point field and overlapping
grains are also called {\em Boolean models}. For these models it is possible
to calculate the volume fraction $\chi_{\rm A}$ analytically
\cite{stoyan83,stoyan87,hermann91,porod52}: 
\begin{equation}
\label{volpoi}
\chi_{\rm A} = 1 - \exp [ -\lambda \bar{V}(A) ] \pu
\end{equation}
Here, $\lambda$ is the number density of the Poisson point field and
$\bar{V}(A)$ is the mean volume of the grains. Equation (\ref{volpoi})
is equivalent to equations (\ref{JMA2} - \ref{JMA4}) if the
corresponding nucleation and growth laws are inserted.\\
To model a system with increased nucleation probability in certain
regions, we use a {\em cluster point field}. The nuclei are  distributed
uniformly and 
independently within spheres of radius $R_{\rm cl}$. These spheres
are distributed according to a Poisson point field with parameter
$\lambda_{\rm par}$, the numbers of points within the spheres are
Poisson distributed with mean value $N_{\rm cl}$. Hence, the density of
the cluster point field is $\lambda_{\rm cl} = \lambda_{\rm par} N_{\rm
cl}$. \\
An appropriate set of parameters for characterizing the cluster model is
$(\lambda_{\rm cl}, N_{\rm cl}, c)$ where $c = 2 R_{\rm cl} / \bar{r}_1$.
Here, $\bar{r}_1$ is the mean distance of the midpoints of neighbouring clusters
given by \cite{hermann91}:
\begin{equation}
\bar{r}_1 = \left ( \frac{3}{4 \pi} \right )^{1/3} \Gamma(4/3) \lpar
^{-1/3} \approx 0.554 \lpar ^{-1/3} \pu
\end{equation}
For $c \gg 1$ the point field approaches a Poisson point field of
density $\lambda_{\rm cl}$.\\
A {\em hard-core point field} is used to force a certain minimum distance
between the nuclei. The distance between any points of the model with density
$\lhc$ is forbidden
to be smaller than a given value $R_{\rm hc}$. The essential property of
the structure is described by the packing fraction $\phc = 4 \pi
/3 R_{\rm hc}^3 \lhc$. For $\phc \rightarrow 0$ the point field
approaches a Poisson point field.\\
Neither for the cluster model nor for the hard-core model analytical
expressions are known that describe the volume fractions.\\
The small-angle scattering intensity $I(q)$ per unit volume is given by
\cite{sonntag81}: 
\begin{equation}
\label{sans}
I(q) = 4 \pi \int\limits_0^{\infty} r^2 [ C_{\rm A}(r) - \chi^2_{\rm A}(r) ]
\frac{\sin(qr)}{qr} \d r \pu
\end{equation}
To calculate the small-angle scattering intensities of the germ-grain
models, the covariance $C_{\rm A}(r)$ of the systems is
needed. This is the 
probability $P$ of two random points $\vec{r}_1, \vec{r}_2$ with distance
$r$ both lying in the region covered by the model:
\begin{equation}
\label{cov}
C_{\rm A}(r) = P(\vec{r}_1 \in A, \vec{r}_2 \in A)\:,\: r = \mid \vec{r}_1 -
\vec{r}_2 \mid \pu 
\end{equation}
For Boolean models an analytical expression \cite{porod52} exists:
\begin{equation}
\label{cov-poi}
C_{\rm A} = 2 \chi_{\rm A} - 1 + (1 - \chi_{\rm A})^2 \exp \left [\lambda
\gamma ^0 (r) \right ]
\end{equation}
where $\gamma ^0 (r)$ is the mean distance probability function averaged
over all spheres with density $f(x)$ of the radii distribution:
\begin{equation}
\label{gamma}
\gamma ^0 (r) = \frac{4 \pi}{3} \int\limits_{r/2}^{\infty} x^3 \left ( 1 -
\frac{3r}{4x} + \frac{r^3}{16 x^3} \right ) f(x) \d x \pu
\end{equation}
The constructional details of the point fields mentioned above are
explained in the next section.

\section{\label{simulation}The simulation technique}
The nucleation and growth processes are simulated in a cube of unit
lenght $L_0$ and volume $V_0 = L_0^3$. All lengths are scaled to
$L_0$. To model an infinite 
structure, periodic boundary conditions are applied. We use
spheres of equal (instantaneous nucleation) or different (continuous
nucleation) size as grains. They grow according to the specified growth law and
nuclei are generated randomly according to the underlying point field. In the
case of instantaneous nucleation (INST), all grains start growing at
$t=0$. If continuous nucleation (CONT) is considered, in every
evolutionary step $\d t$ a mean number of nuclei starts
growing according to 
the nucleation rate $\alpha(t)$. In this case, the nuclei that are
created in an already transformed area have to be ommitted.\\
After every time step, the volume fraction $\chi_{\rm A}$ of the system
is calculated. Optionally, the covariance $C_{\rm A}(r)$ of the
structure can be calculated at a given volume fraction.\\
In every simulation, 500 - 700 nuclei are generated. To limit the
influence of statistical fluctuations, the whole procedure is repeated
10 - 40 times and the average of the relevant quantities is evaluated.

\subsection{Construction of the point fields}
In a first step of the simulation the underlying point field has to be
created.\\
To generate a Poisson point field, the number $N_{\rm nu}$ of nuclei is
drawn from a Poisson distribution with parameter $\lambda V_0$. Then
$N_{\rm nu}$ points with 
cartesian coordinates equally distributed in $(0, L_0)$ are created.\\
A cluster point field is build by a two-step procedure. First, a
Poisson point field of parent points with parameter $\lambda_{\rm par}$
is created in $V_0$. In a second step, spheres of radius $\rcl$ are
attached to these parent points. In each of these spheres, another
Poisson process with parameter $\ncl$ is created. Omitting the parent
points, the remaining nuclei obey a cluster distribution with density
$\lcl = \lambda_{\rm par} \ncl$.\\
To construct a hardcore point field, single points with equally
distributed coordinates are generated subsequently. New points are
accepted only if their 
distance to all existing points is greater than $2\rhc$. This procedure is
repeteated until the desired number of nuclei (Poisson deviated with
parameter $\lhc V_0$) is reached.\\

\subsection{Nucleation and growth}
After the definition of the point fields, the time evolution of the
system starts.\\
In the case INST, all predefined grains start growing at $t=0$. In every
time step, their radii are calculated as $r(t) = \int_0^t v(\tau) \d
\tau$. All grains have the same size.\\
If continuous nucleation is simulated, in every time step $N_{\rm
act}$ nuclei start growing. $N_{\rm act}$ is Poisson deviated with the
parameter $\alpha (t) \d t$. The radii of the grains are different now,
depending on the life time $t_i$ of each individual grain $i$:
\begin{equation}
r_i(t_i) = \int\limits_0^{t_i} v(\tau) \d\tau \pu
\end{equation}

\subsection{Calculation of the volume fraction and the covariance}
To calculate the volume fraction, 
a fine grid of $N_{\rm test}$ test points is constructed in the unit
cube. The coordinates of 
these test points are determined using a quasi-random sequence according
to Sobol \cite{sobol67,antonov79,nrc}. The volume fraction is given
by the number of test points $N_{\rm in}$ that lie within the area
covered by the spheres:
\begin{equation}
\label{volsim}
\chia = \frac{N_{\rm in}}{N_{\rm test}} \pu
\end{equation}
To evaluate the covariance, the distances between all the
test points are calculated. The numbers $N(r)$ of test points that lie
within discrete distance intervalls between $r$ and $r + \d r$ are
determined. Then, the covariance is given as
\begin{equation}
\label{covsim}
C_{\rm A} (r) = \frac{N_{\rm in}(r)}{N(r)} 
\end{equation}
where $N_{\rm in}(r)$ denotes the number of test points in the
distance intervall $(r,r  + \d r)$ that are covered by a sphere.\\
The small-angle scattering intensities can now be calculated according
to equation (\ref{sans}). To solve the integral, we use a Fourier transfom
method and fit a polynomial to the discrete values of $C_{\rm A} (r) -
\chia ^2$ in order to obtain the function to be integrated.

\subsection{Test of accuracy}
\begin{figure}[tbh]
\epsfxsize=\textwidth
\epsffile{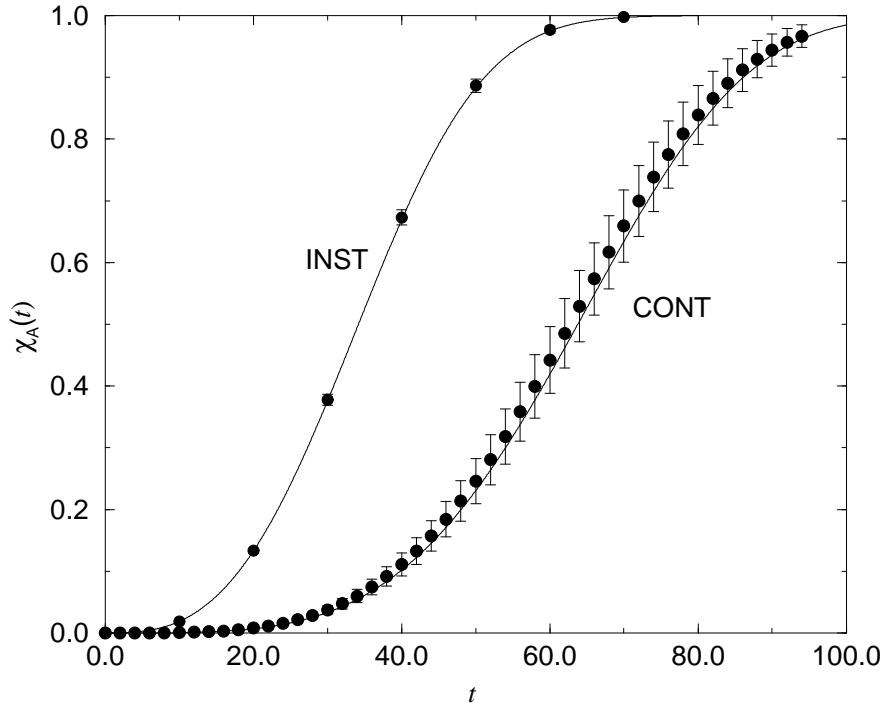}
\caption{\label{testvol1}Comparison of the simulated volume fractions
($\bullet$) with the exact JMA-equation (--) in the case of instantaneous
nucleation (INST) and homogeneous nucleation (CONT). The error bars denote
the standard deviation of the results. INST: $\lambda=500$, $v = 0.002$,
CONT: $\alpha=5$, $v=0.002$, $\d t = 1$.}  
\end{figure}

\begin{figure}[tbh]
\epsfxsize=\textwidth
\epsffile{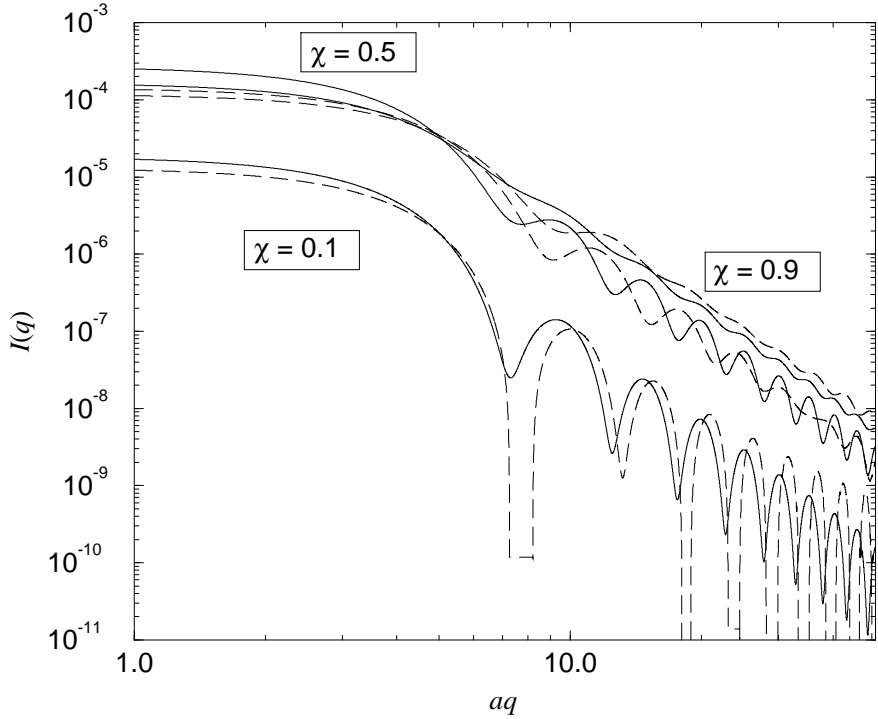}
\caption{\label{sanstest}Comparison of the simulated small-angle
scattering values (dashed lines) with exact results (solid lines) for
Poisson distributed nuclei at
different volume fractions. The parameter $a$ denotes the radius of the
grains. $\lambda = 500$.}
\end{figure}

\begin{figure}[tbh]
\epsfxsize=\textwidth
\epsffile{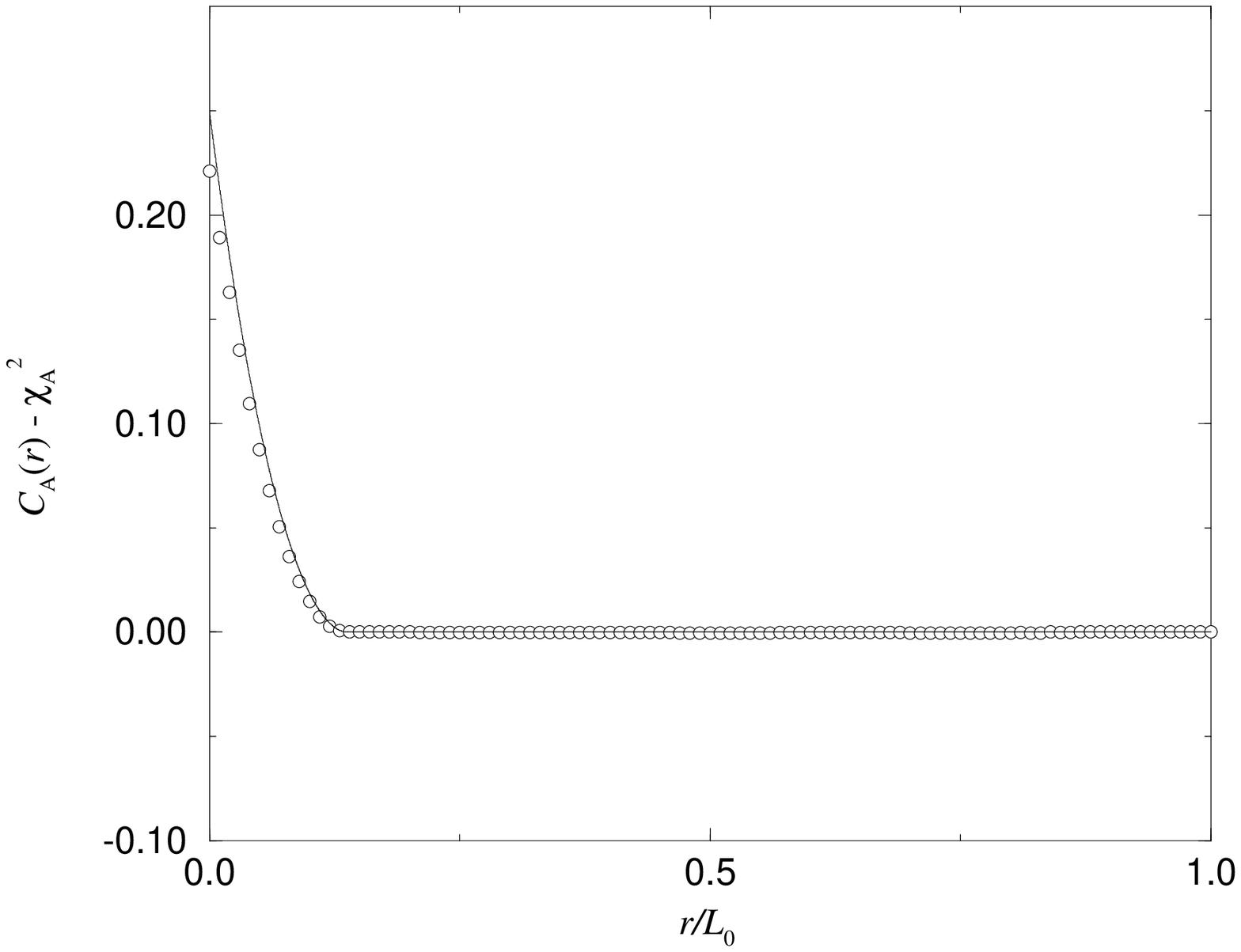}
\caption{\label{covtest}Comparison of the simulated covariance values
($\circ$) with exact results (--) according to
(\ref{cov-poi}). Parameters as in \fref{testvol1} (INST).}
\end{figure}

Due to the finite and discrete nature of the simulation, two sources of
systematic errors have to be considered:
\begin{itemize}
\item The size of the system (i.e.~the number of grains) has to be large
enough in order to describe an infinite system.
\item In the simulation of CONT, the nucleation proceeds in small but
finite time steps $\d t$.
\end{itemize}
In order to check the accuracy of the simulation, we performed
calculations using the Poisson point field and compared the results for
the volume fraction and the covariance with the exact equations
(\ref{JMA2}) and (\ref{JMA3}), respectively.\\
In figure \ref{testvol1}, the results of simulations in the case of
instantaneous and continuous nucleation are shown. The simulations were
performed 10 times with a step size $\d t = 10$ (INST) resp. $\d t = 1$
(CONT)\footnote{The time scaling does not have any influence on the
resulting Avrami exponent.} and $5\cdot 10^4$ test points were used. An
Avrami analysis by 
means of linear regression yielded an Avrami exponent $n = 2.96$ for the
simulation of instantaneous nucleation, which is in good agreement with
the exact value $n = 3$. \\ 
In the case of continuous nucleation, the value of the Avrami exponent is
$n=3.95$, compared with the exact value $n=4$.\\
The simulated small-angle scattering intensities of
instantaneous 
nucleation (parameters as above) at different volume fractions are shown
in figure 
\ref{sanstest}. Here, quantitative differences between the
simulated values and the exact ones calculated according to
(\ref{sans}), (\ref{cov-poi}) and (\ref{gamma}) occur, although the covariance
values are in quite good agreement, see figure \ref{covtest}. In both
cases, the same numerical integration method was used. Because
of the multiplication of $\cova (r) - \chia ^2$ by $r$ in equation
(\ref{sans}), very small differences between the simulated covariance values
and the exact ones at large r-values yield substantial differences
after integration.\\ 
Hence, with the present accuracy of our method, only qualitative
statements concerning the small-angle scattering curves are
possible. But the main features of the scattering intensities at
different volume fractions are represented properly. The curve with low
volume fraction shows well-resolved maxima and minima as the single
spheres are still nearly isolated. With increasing volume fraction, the
amplitudes of the oscillations decrease, at $\chia = 0.9$ there are only
weak ripples left.

\section{\label{results}Results and discussion}
We investigated the dependence of crystallization kinetics on the spatial
distribution of the nuclei in the case of continuous
nucleation (CONT) and instantaneous nucleation (INST). Additionally, for
CONT the case of diffusion controlled growth was surveyed. The small-angle
scattering intensities were calculated for instantaneous nucleation in
order to check if it is possible to distinguish between the several
grain distributions by means of small-angle scattering methods.

\subsection{Instantaneous nucleation}
In the case of instantaneous nucleation, we checked the influence of
cluster- and hardcore-model on crystallization kinetics. The results were
compared with the corresponding simulated values for the Poisson
model. We used constant growth rates $v$ for calculations\footnote{Note
that in the case of instantaneous nucleation time dependent growth rates
can be reduced to this case as all grains start growing at the same
time. See \cite{shepilov94}.}.\\
Furthermore, the covariance values and the resulting small-angle
scattering intensities were evaluated.

\subsubsection{Results on crystallization kinetics}

\begin{figure}[tbh]
\epsfxsize=\textwidth
\epsffile{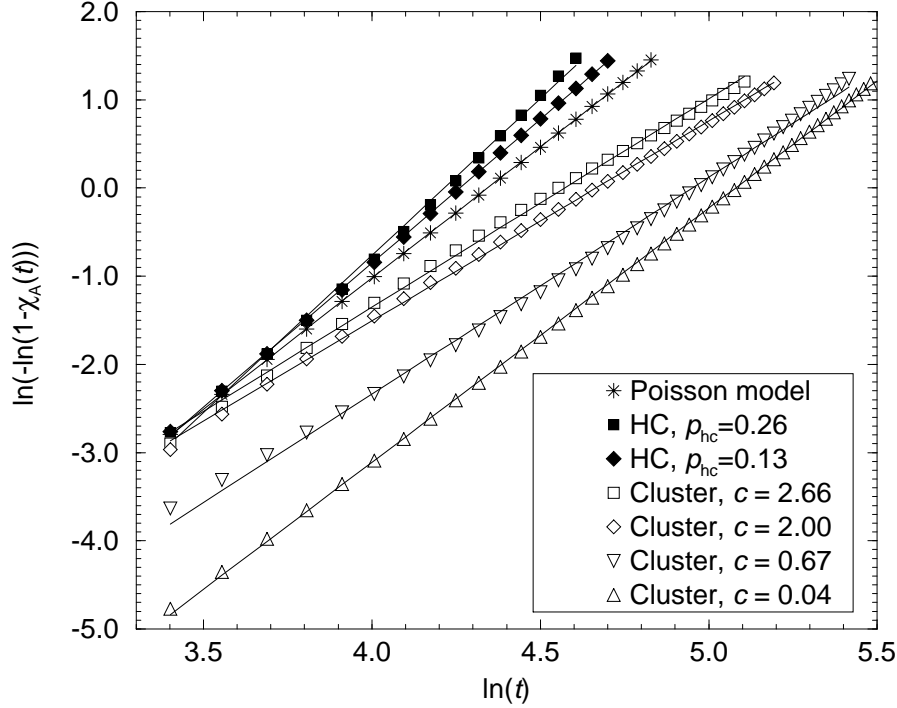}
\caption{\label{avr-cluhc-nonu}Influence of the parameter $c$ of the
cluster model and the parameter $\phc$ of the hardcore model on the Avrami
exponent in case INST. The straight lines are linear 
regression values. $\lambda = 500, N_{\rm cl} = 10$.}
\end{figure}

In figure \ref{avr-cluhc-nonu}, Avrami plots of a cluster model and a
hardcore model are
compared with that of a Poisson model with equal point density
$\lambda$. It is clearly shown that the distribution of the nuclei
according to a cluster model leads to a reduced Avrami exponent $n$
compared with the value $n=3$ expected for the Poisson
model. For small values of $c$, the simulated values deviate from a
straight line in the Avrami plot, which means that the crystallization
kinetics cannot be represented by an exponential law according to
\eref{JMA4} in these cases. If $c$ is small enough, $n$ raises again and
the deviations from the exponential law decrease again.
The values of the simulated Avrami exponents (drawn from a linear
regression analysis) are shown in table \ref{avr-cluhc-nonu.tab} for two
simulations with
($\lcl = 50, \ncl = 10, c$) and ($\lcl = 25, \ncl = 20, c$), respectively.
\begin{table}
\caption{\label{avr-cluhc-nonu.tab}Avrami exponents for several
parameter values of the cluster distribution and the hardcore model in case
INST with $\lambda = 500$. Poisson model: $n = 2.97$} 
\begin{indented}
\item[]\begin{tabular}{@{}llllll}
\br
\multicolumn{2}{@{}l}{Cluster, $\ncl = 10$} &
\multicolumn{2}{l}{Cluster, $\ncl = 20$} &
\multicolumn{2}{l}{Hardcore}\\
$c$ & $n$ & $c$ & $n$ & $\phc$ & $n$\\
\mr
2.66 & 2.36 & 2.11 & 2.08 & 0.26 & 3.56\\
2.00 & 2.26 & 1.58 & 2.01 & 0.13 & 3.27\\
1.33 & 2.26 & 1.06 & 2.07 & 0.06 & 3.07\\
0.67 & 2.46 & 0.53 & 2.33 & 0.02 & 2.99\\
0.33 & 2.43 &      &      &      &     \\
0.17 & 2.64 &      &      &      &     \\
0.08 & 2.80 &      &      &      &     \\
0.04 & 2.88 &      &      &      &     \\
\br
\end{tabular}
\end{indented}
\end{table}
Simulation results in the case of a hardcore model with $\lhc = 500$ and
several hardcore radii $\rhc$ are also listed. The simulation of the
corresponding JMA case yields an Avrami exponent $n = 2.97$. 
According to these simulation results, a 
distribution of the nuclei according to a hardcore model leads to an
increased Avrami exponent $n$ compared with the one expected by the JMA
theory. A deviation from the linear behaviour in the Avrami plot can
also be observed for large values of $\phc$.\\

\subsubsection{Results on small-angle scattering data}
\begin{figure}[tbh]
\epsfxsize=\textwidth
\epsffile{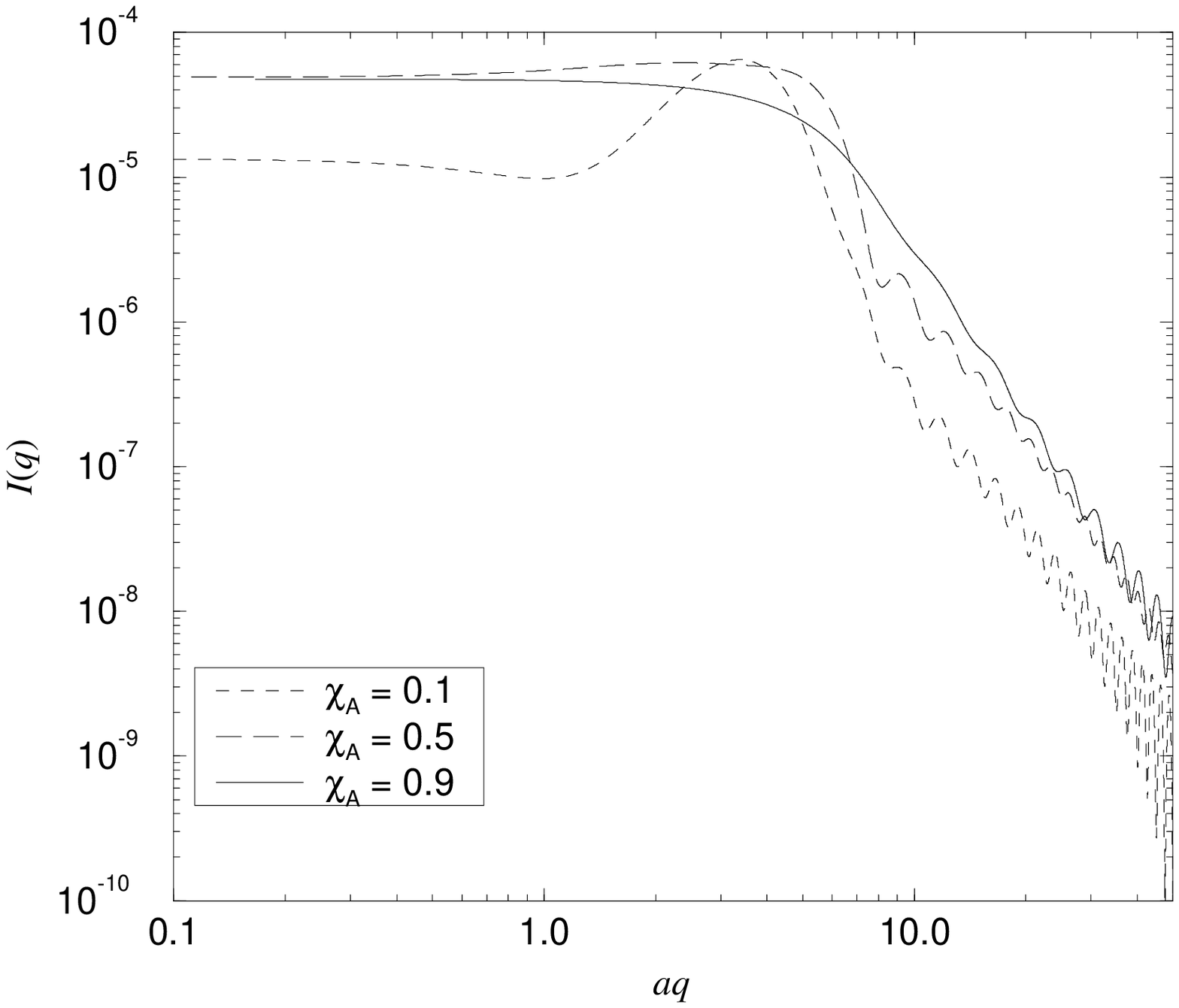}
\caption{\label{sanshc}Simulated small-angle scattering curve of a harcore model
with different volume fractions. $\lhc = 500$, $\rhc = 0.045$. The
parameter $a$ denotes the radius of the grains.}
\end{figure}

\begin{figure}[tbh]
\epsfxsize=\textwidth
\epsffile{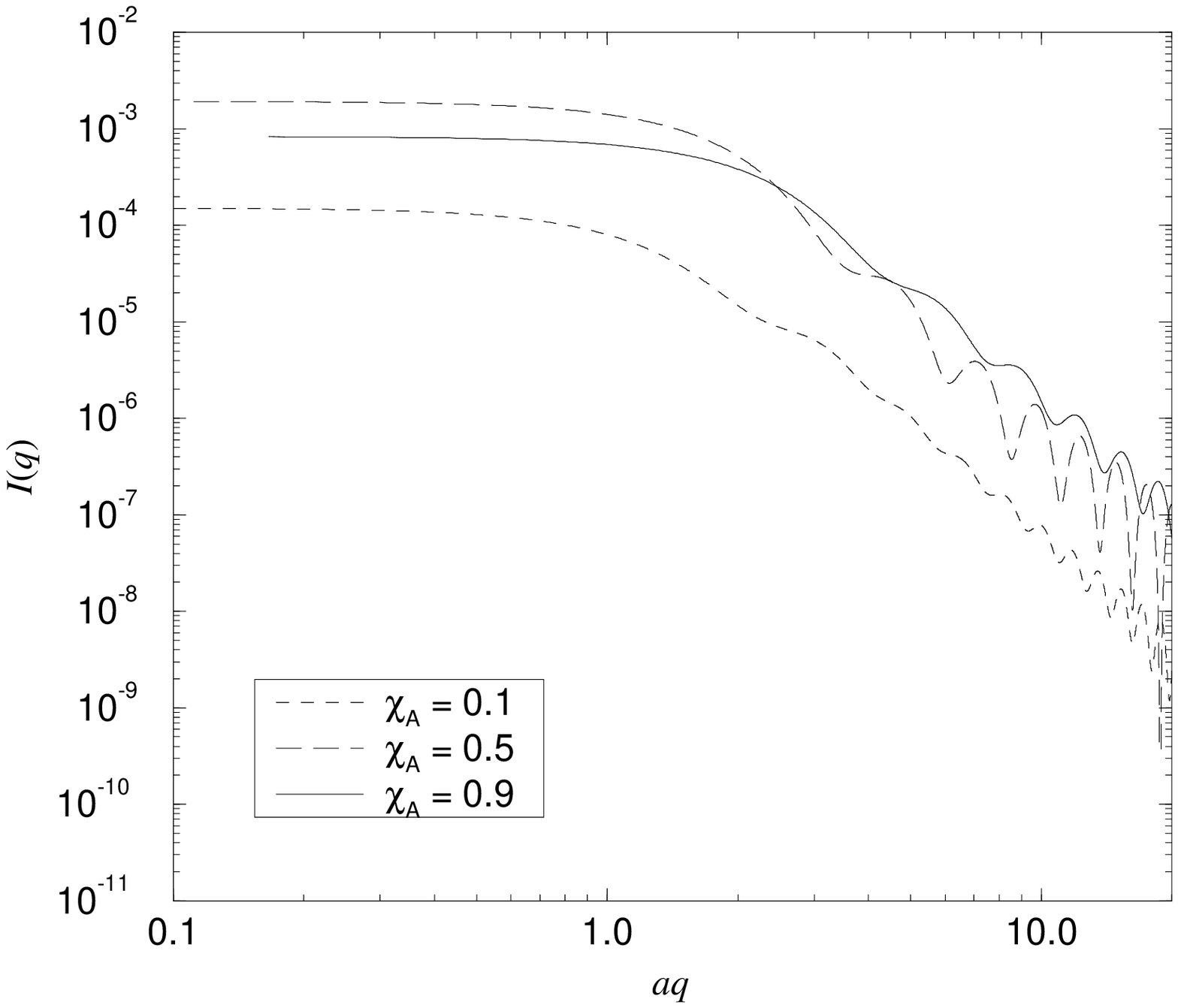}
\caption{\label{sansclu}Simulated small-angle scattering curve of a
cluster model with different volume fractions. $\lcl = 50$, $\ncl = 10$,
$c = 2.0$. The parameter $a$ denotes the radius of the grains.}
\end{figure}

To check the possibility of distinguishing experimentally between the
distributions under consideration, we investigated the small-angle
scattering curves at several volume fractions.\\
The scattering curves of the Poisson model were already shown in
\fref{sanstest}. \Fref{sanshc} shows scattering curves of a germ-grain
model with underlying hardcore distribution. For low volume fractions,
the curves exhibit a significant first peak. This peak is characteristic
for a hard-sphere model with non-overlapping spheres (see,
e.g.~\cite{hermann91}). With increasing volume fraction this peak
disappears as the structure of the system is now far away from the
structure of the generating nuclei and the overlapping of the grains
becomes larger. The scattering curves of a cluster
model with different volume fractions are shown in \fref{sansclu}. Here,
at low volume fractions no sharp peaks are present. With increasing
volume fraction, the amplitude of the oscillations first increases and then
decreases again.\\
Considering these results, it should be possible to distinguish between
the three distributions of the nuclei by using small-angle scattering and
observing the whole crystallization process.

\subsection{Continuous nucleation}
\begin{figure}[tbh]
\epsfxsize=\textwidth
\epsffile{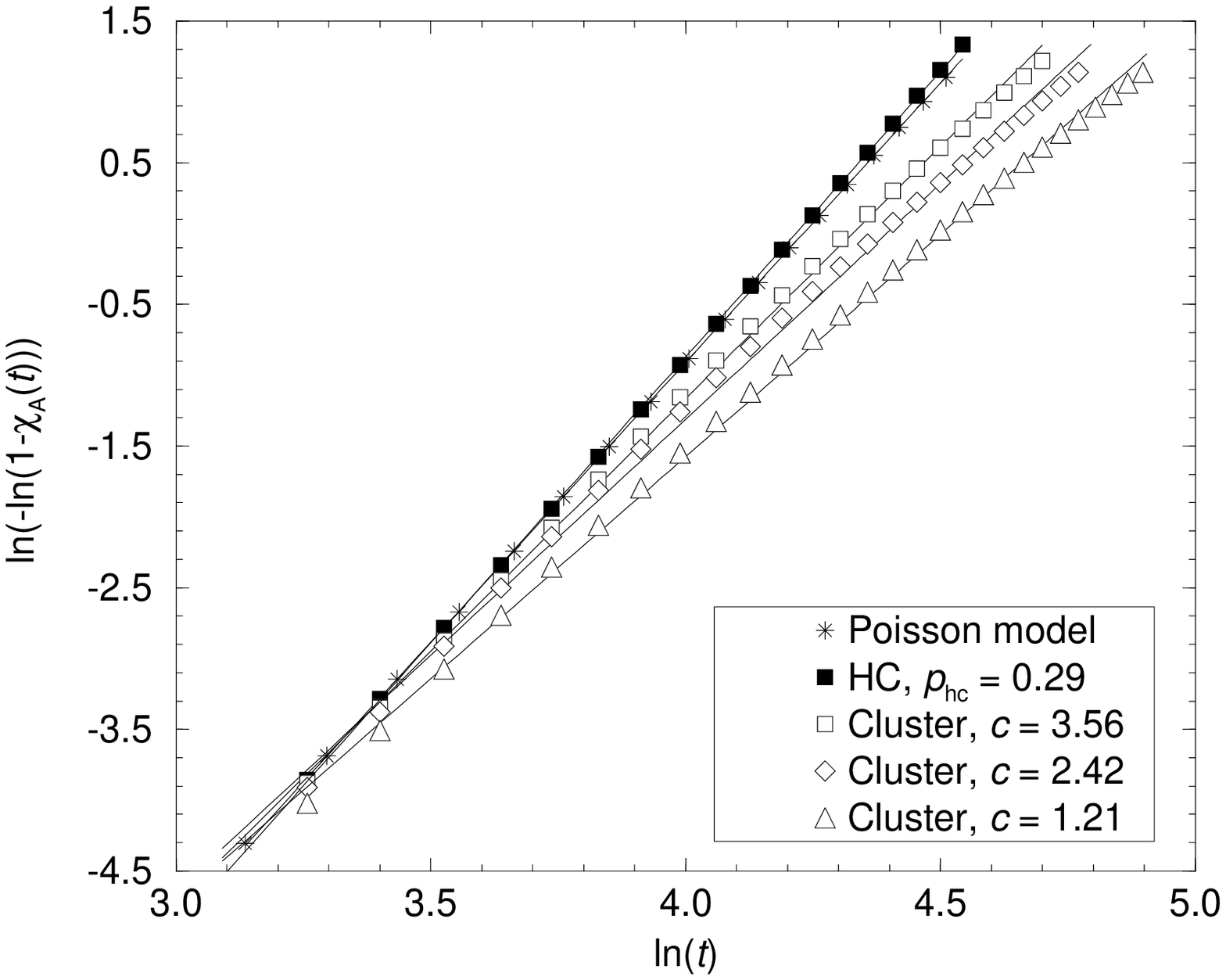}
\caption{\label{avr-cluhc-nu}Influence of the parameter $c$ of the
cluster model and the parameter $\phc$ of a hardcore model on the Avrami
exponent in case CONT. The straight lines are linear 
regression values. $\lambda = 750, \ncl = 20, \alpha = 5 / \d t$.}
\end{figure}

The same investigations as in the case of instantaneous nucleation were
also made for continuous nucleation and linear growth rate $v
=\mbox{const.}$. In the case of diffusion controlled growth, we analysed
the error that is made by applying the JMA equation.\\
In both cases,
predefined point fields of density $\lambda = 750$ and a nucelation
rate $\alpha = 5 / \d t$ were used for the simulation.\\
In \fref{avr-cluhc-nu}, the influence of a cluster and a hardcore point
distribution on crystallization kinetics is shown (the given parameters
describing the point fields refer to the predefined nuclei). Further
simulation results are listed in \tref{avr-cluhc-nu.tab}. 
\begin{table}
\caption{\label{avr-cluhc-nu.tab}Avrami exponents for several
parameter values of the cluster distribution and the hardcore model in case
CONT with $\alpha = 5 / \d t$. Poisson model: $n = 3.95$} 
\begin{indented}
\item[]\begin{tabular}{@{}llllll}
\br
\multicolumn{2}{@{}l}{Cluster, $\ncl = 10$} &
\multicolumn{2}{l}{Cluster, $\ncl = 20$} &
\multicolumn{2}{l}{Hardcore}\\
$c$ & $n$ & $c$ & $n$ & $\phc$ & $n$\\
\mr
4.57 & 3.72 & 3.63 & 3.56 & 0.27 & 4.04\\
3.05 & 3.59 & 2.42 & 3.33 & 0.20 & 3.99\\
1.52 & 3.40 & 1.21 & 3.14 &  & \\
\br
\end{tabular}
\end{indented}
\end{table}
As in the case of INST, a nuclei distribution according to the cluster
model results in a decreased Avrami exponent compared with the JMA
case and in deviations from the exponential behaviour. A hardcore
distribution leads to an increased Avrami exponent, but the differences
to the Poisson model are not as distinct as in case INST.

\subsection{Discussion of crystallization kinetics}
The deviations from the Poisson model in the cases INST and CONT can be
explained by looking at the derivation of the JMA equation (see,
e.g.~\cite{christian}). To calculate the transformed volume fraction,
a so-called {\em extended volume} is introduced, which is simply the sum
of the volumes of all grains without considering their mutual overlappings.
To get a connection between this extended volume and the real
transformed volume, assumption (\ref{2}) is applied.\\
If the grains are distributed according to the cluster model, their
overlapping is underestimated by the JMA model and hence the transformed
volume fraction is overestimated. For
$c \gg 1$ (clusters lose their cluster-like nature), $n$ is close to the
JMA-value. With decreasing $c$ of the underlying cluster point field,
the Avrami exponent $n$ gets smaller and substantial deviations from the
exponential behaviour of the JMA-kinetics occur. If $c \rightarrow 0$,
the Avrami exponent increases again and the Avrami plot shows a linear
behaviour. In this case, the clusters are widely spaced and act
like single Poisson-distributed nuclei. In between these two limiting
cases ($c \gg 1$ and $c \rightarrow 0$), deviations from the
JMA-behaviour occur. Values of $c$ with approximately $1 < c < 2$ give
the maximum deviation (minimum $n$).\\
On the other side, a hardcore distribution leads to a
smaller overlapping compared with a uniform distribution. The transformed
volume fraction is larger than in the pure JMA case, since more of the
space nuclei grow into is empty. Our simulations show an increase of the
Avrami exponent $n$ with increasing packing fraction $\phc$ of the
underlying point field. On the other hand, in the case of $\phc
\rightarrow 0$ $n$ reaches the value of the pure
JMA-case. Unfortunately, with our present algorithm we could not reach
the limiting case of a close-packing of the underlying point field.\\
As outlined in \sref{intro}, in the case of diffusion controlled growth
the crystallized volume fraction is overestimated. To check for this, we
performed simulations with an underlying Poisson distribution of the
nuclei. In a first series, the phantom nucleii that nucleated in an
already crystallized region were discarded. Afterwards, they were
treated like regular grains and contributed to the volume
fraction. Doing so, we could estimate the
error that is made in applying the JMA equation on diffusion controlled
growth. The calculation of the differences $\Delta _{\rm abs}$ of the volume
fractions yielded differences $\Delta _{\rm abs} \leq 9 \cdot 10^{-3}$. 
These results are in good agreement with simulations made by Shepilov
and Bochkarev \cite{shepilov87}.\\

\section{Conclusions}
Our simulations concernig the dependence of crystallization kinetics on the
spatial nuclei distribution clearly showed that the analysis of
experimental data by the JMA equation must be done with care. If the
nuclei are not distributed equally, the use of the JMA equation can yield
substantially wrong results. 
Therefore it should be checked if the JMA
equation is applicable. One possibility to do so is the use of
small-angle scattering. The scattering curves of the investigated nuclei
distributions differ clearly from one another, especially in an early
stage of the crystallization process.\\
On the other hand, the simulations showed that the error that is made by
applying the JMA equation on diffusion controlled growth processes with
continuous nucleation can be neglected.\\

\end{document}